\documentclass[prl,letter,twocolumn,nopacs,superscriptaddress,notitlepage]{revtex4-2}
\usepackage[utf8]{inputenc}
\usepackage[american,]{babel}
\usepackage[T1]{fontenc}
\usepackage[pdftex]{graphicx}  
\usepackage{graphicx, xcolor}
\usepackage{dcolumn}
\usepackage{bm}
\usepackage{amsmath,amsthm,amssymb}
\usepackage{hyperref}
\usepackage[T1,T2A]{fontenc}
\usepackage{xcolor}
\hypersetup{colorlinks,bookmarksopen,bookmarksnumbered,
    citecolor=blue,
    linkcolor=blue,
    pdfstartview=false,
    urlcolor=blue}
\usepackage{graphicx}
\usepackage{braket}
\usepackage{wrapfig}
\usepackage{soul}
\usepackage{mathtools}
\usepackage{float}
\usepackage{tabularx}
\usepackage{tikz}
\usepackage{dsfont}
\usepackage{subcaption}
\begin{document}

\newcommand{\titleinfo}{Measurement-induced phase transitions by matrix product states scaling}
\title{\titleinfo}

\author{Guillaume Cecile}
\affiliation{Laboratoire de Physique Th\'eorique et Mod\'elisation, CNRS UMR 8089,
CY Cergy Paris Universit\'e, 95302 Cergy-Pontoise Cedex, France}
\author{Hugo L\'oio}
\affiliation{Laboratoire de Physique Th\'eorique et Mod\'elisation, CNRS UMR 8089,
CY Cergy Paris Universit\'e, 95302 Cergy-Pontoise Cedex, France}

\author{Jacopo De Nardis}
\affiliation{Laboratoire de Physique Th\'eorique et Mod\'elisation, CNRS UMR 8089,
CY Cergy Paris Universit\'e, 95302 Cergy-Pontoise Cedex, France}

\begin{abstract}
We study the time evolution of long quantum spin chains subjected to continuous monitoring via matrix product states (MPS) at fixed bond dimension, with the Time-Dependent Variational Principle (TDVP) algorithm. The latter gives an effective classical \textit{non-linear} evolution with a conserved charge, which approximates the real quantum evolution up to an error. We show that the error rate displays a phase transition in the monitoring strength, which can be well detected by scaling analysis with relatively low values of bond dimensions. The method allows for an efficient numerical determination of the critical measurement-induced phase transition parameters in many-body quantum systems.  Moreover, in the presence of $U(1)$ global spin charge, we show the existence of a charge-sharpening transition well separated from the entanglement transition which we detect by studying the charge fluctuations of a local sub-part of the system at very large times.
Our work substantiates the TDVP time evolution as a method to identify measured-induced phase transitions in systems of arbitrary dimensions and sizes.
\end{abstract}

\maketitle
\textit{Introduction.---}
The dynamics of quantum many-body systems are particularly challenging due to the rapid growth of entanglement throughout the system. Indeed, even if the initial state is a simple product state, unitary time evolution typically entangles the different degrees of freedom in the system \cite{doi:10.1080/00018732.2016.1198134,PhysRevX.7.031016}, and the wave function rapidly becomes exponentially complex.  This picture is challenged when on top of the unitary dynamics one adds an externally induced non-hermitian contribution to the evolution, for example when the system is continuously monitored or measured at a given rate $\gamma$ \cite{li2019measurementdrivenentanglement,chan2018unitary, PhysRevX.9.031009,vijay2012stabilizing,katz2006coherent,campagneibarcq2016observing,bauer2015computing,PhysRevB.106.L220304}.  In this case, the average von Neumann entanglement entropy of the system transits from a volume law scaling to an area law, depending on the rate of monitoring. Such a phase transition has been dubbed measured-induced phase transition (MIPT) and has received a large experimental and theoretical attention in the past years due to its numerous and inspiring connections with quantum computing and error correction theory, for example ~\cite{PhysRevX.9.031009,nahum2921measurementandentanglement,jian2020measurementinducedcriticality,10.21468/SciPostPhys.7.2.024,potter2022entanglementdynamicsin,chan2019unitaryprojective,li2019measurementdrivenentanglement,skinner2019measurementinducedphase,czischek2021simulating,han2022entanglementstructure,minoguchi2022continuousgaussianmeasurements,altland2022dynamicsofmeasured,fuji2020measurementinducedquantum,jian2021yangleeedge,bentsen2021measurementinducedpurification,yang2022entanglementphasetransitions,medina2021entanglementtransitionsfrom,lunt2020measurementinducedentanglement,szyniszewski2019entanglementtransitionfrom,tang2020measurementinducedphase,iadecola2022dynamicalentanglementtransition,odea2022entanglementandabsorbing,ravindranath2022entanglementsteeringin,PhysRevLett.130.120402,sierant2023entanglement,buchhold2022revealing,vijay2020measurementdrivenphase,fan2021selforganizederror,li2021conformal,ippoliti2021entanglementphasetransitions,klocke2022topologicalorderand,PhysRevX.13.041028,li2021robustdecodingin,li2021statisticalmechanicsmodel,li2021entanglementdomainwalls,li2021statisticalmechanicsof,feng2022measurementinducedphase,barratt2022transitions,zabalo2022operatorscalingdimensions,zabalo2020criticalpropertiesof,sierant2022universalbehaviorbeyond,iaconis2020measurementinducedphase,han2022measurementinducedcriticality,liu2022measurementinducedentanglement,sang2021entanglementnegativityat,shi2020entanglementnegativityat,weinstein2022measurementinducedpower,Turkeshi_2022,turkeshi2024density,turkeshi2020measurementinducedcriticality,turkeshi2022measurementinducedcriticality,sierant2022measurementinducedphase,zabalo2022infiniterandomnesscriticality,PhysRevResearch.4.043212,Morral_Yepes_2023,10.21468/SciPostPhys.14.3.031,biella2021manybodyquantumzeno,PhysRevB.103.224210,2023Google,altman2023,PhysRevResearch.4.023146,PhysRevB.107.214203,PhysRevX.13.041046,doggen2023ancilla,poboiko2023measurementinduced}.

Here we aim to detect MIPT as a simulability transition by tensor network wave functions (a slightly related approach can be found also in \cite{Azad2023}). We consider matrix product states (MPS), which are well-known efficient representations of any wave function with area law entanglement scaling, \cite{MatrixProductStateRepresentations,Verstraete2008,PhysRevB.73.094423}: i.e. given the bond dimension of the MPS $\chi$, one expects that an MPS wave function can represent any area law wave functions only up to exponentially small (in bond dimension $\chi$) corrections.  Doing time evolution with MPS methods is instead challenging, as the spread of entanglement throughout the system makes their bond dimension grow, typically exponentially with time. However, in the presence of strong monitoring, the bond dimension is expected to saturate at large times, signalling this way the emerging area law phase. 

 Partially motivated by ideas of hydrodynamics, where classical non-linear dynamics with the necessary conservation laws well reproduce large-scale proprieties, we here consider the time evolution of MPS with fixed bond dimension as given by the Time-Dependent Variational Principle (TDVP) ~\cite{PhysRevB.94.165116,PhysRevB.88.075133,leviatan2017quantum,PhysRevB.97.024307,NatureComms41467-019-10336-4,Hallam2019,Azad2023,PhysRevB.101.081411,PhysRevResearch.4.023146}. Here, the Hamiltonian time evolution at each time step is projected back on the manifold of MPS with fixed bond dimensions $\chi$. Such evolution is known to conserve total energy and any other local conserved charge, such as total magnetization. Moreover, the TDVP time evolution, as the projection is state-dependent, is a \textit{non-linear} evolution, as opposed to other methods for MPS evolution.  Therefore, the average of monitored projected trajectories is expected to be much different from the average state. Here we show that finite-$\chi$ scalings of the transition are very well present in this effective projected quantum evolution.  
When projecting the unitary time evolution of a Hamiltonian into the space of MPS with bond dimension $\chi$, the norm of the orthogonal vector to the space at any time step gives a measure of the distance between the true quantum evolution and the projected one.  As the maximal entanglement entropy contained in an MPS is $\log \chi$, such \textit{error rate} decays as $1 /\log \chi$ for unitary evolution. We here show that this scaling changes drastically as the monitoring rate is increased, transmuting to exponential decay, for $ \gamma > \gamma_c$. We characterize this transition, which we conjecture to be the same as the entanglement transition, by employing a rescaling in bond dimension $\chi$ using relatively small values and with no limits in the system size, accessing, therefore, the MIPT in regimes which would be impossible for other numerical methods. Analogously to the equilibrium case, see in particular \cite{PhysRevLett.102.255701,PhysRevB.78.024410,PhysRevLett.129.200601,PhysRevLett.123.250604}, where the finite bond dimension introduces an effective length $L(\chi) \sim \chi^\kappa$, and where scaling in bond dimension can be usefully used to probe critical phenomena,  our approach extends these ideas to the study of MIPTs.

Moreover, given that the TDVP evolution preserves the eventual global $U(1)$ symmetry of the Hamiltonian, we employ a scaling to study the charge-sharpening (CS) transition, which is conjectured to occur at smaller measurement rates compared to the entanglement transition \cite{agrawal2022entanglmentandchargesharpening,vasseurexperiment,PhysRevLett.129.120604,PhysRevB.107.014308}. In order to extend the CS protocol to large system sizes, we here consider the variance of the fluctuations of the local magnetization on a sub-system of size $\ell$, e.g. by defining $Q_\ell = \sum_{j \in \ell} S^z_j $, we introduce its variance as 
\begin{equation}\label{eq:Well}
   W^2_\ell =  \langle Q^2_\ell \rangle - \langle Q_\ell \rangle^2 = \sum_{i,j \in \ell} \langle S^z_i S^z_j \rangle^c \ , 
\end{equation}
averaged over quantum trajectories (as for all the other quantities we compute), and we show that the latter shows a transition at large times from an ETH-like extensive behaviour $W^2_\ell \sim \ell$ \cite{https://doi.org/10.48550/arxiv.2306.11457,PhysRevLett.131.210402,PhysRevB.107.014308} to a sub-extensive one $W^2_\ell \sim \ell^\alpha $ with $\alpha<1$, for $\gamma \geq \gamma_{\#}$. We stress that a sub-extensive variance signals a form of anti-correlated connected spin-spin correlations and one can also find sub-extensive fluctuations with purely unitary evolution in some particular cases \cite{Cecile2023} or in Luttinger liquid ground states, where $W^2_\ell \sim \ell^0$. We find that for any given $\chi \gtrsim 16$ the data can be rescaled well on a piece-wise function such that $f(\gamma - \gamma_\#)$ for $\gamma<\gamma_\#$ and $f(\log \ell (\gamma - \gamma_\#))$ for $\gamma > \gamma_\#$, in perfect agreement with the predicted Kosterlitz–Thouless (KT) scaling \cite{PhysRevLett.129.120604}.  As the fitted $\gamma_\#$ is converged already at small values of bond dimension, we find, in all cases studied, that such transition always appears at a smaller rate than the MIPT $\gamma_{\#} < \gamma_c$.  

\begin{figure}[t]

    \centering
    \resizebox{\linewidth}{!}{
    \def\width{4in}
    \begin{tikzpicture}
      \node[draw = none, fill = none] at (0,4.7){\includegraphics[trim={5 5 5 5}, clip, width=\width]{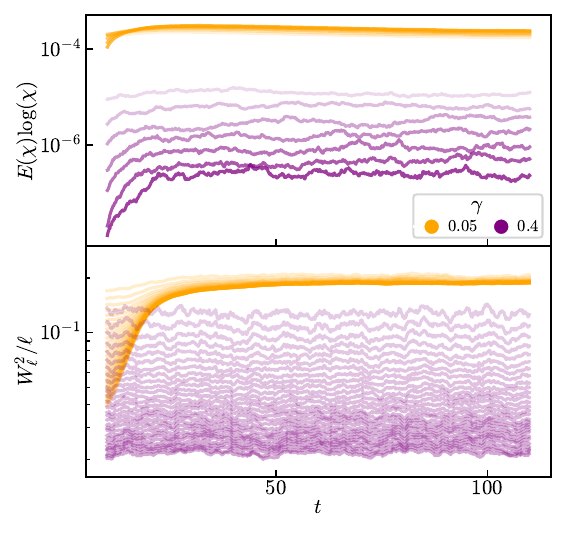}};
      
      
      \node[scale = 1.3] at (-4.8,9.2) {\textbf{(a)}};
      \node[scale = 1.3] at (-4.8,4.9) {\textbf{(b)}};

    \end{tikzpicture}
    }

\caption{\textbf{Monitored XXX chain initialised in the Néel state}: (a) time evolution of the error rate multiplied by $\log \chi$ with TDVP evolution of a chain with $L=60$ sites with different values of $\chi = n^2$, $n\in [4,10]$ (light to dark shades of colour) and with two values of $\gamma$, one in the volume law phase and one in the area law phase.  (b) Same for the charge fluctuations on sub-systems of size $\ell$ with different $\ell \in [2,30]$.  }
\label{fig_time_evol}
\end{figure}

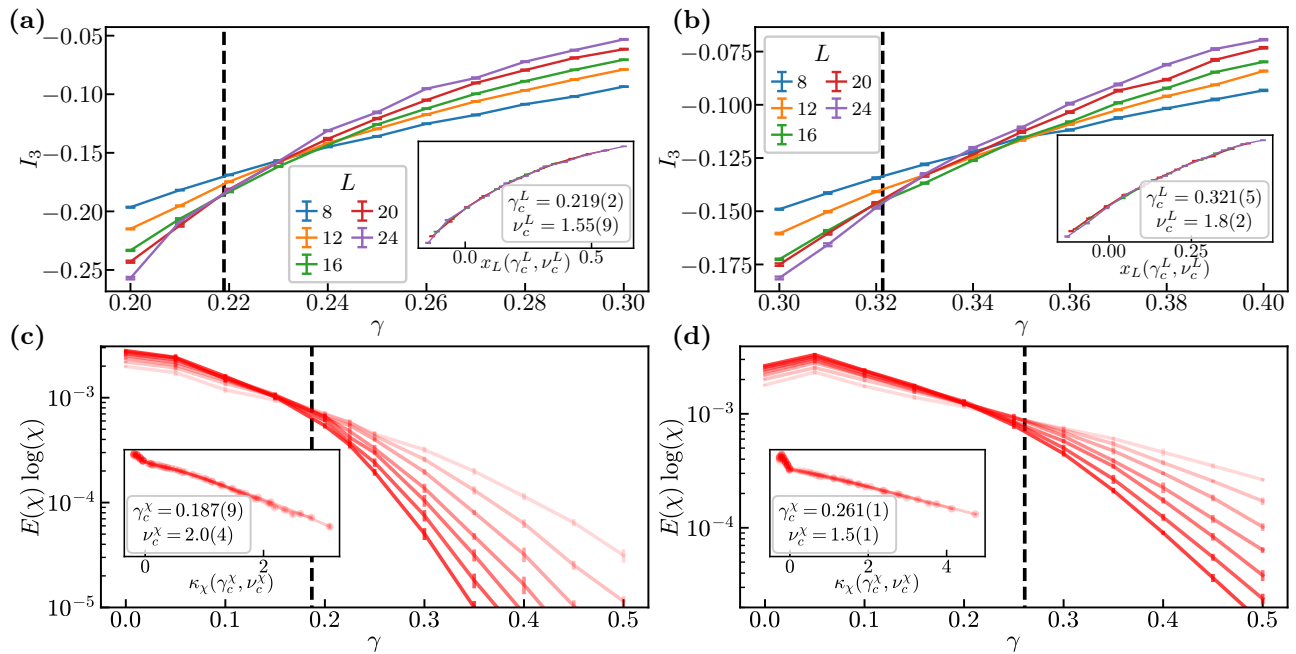
\begin{figure*}[t]
  \centering
  \resizebox{\linewidth}{!}{
    \def\width{4in}
    \def\widthinset{1.4in}
    \def\heightinset{0.66in}
    \def\height{2in}
    \def\scale{1.3}
    \begin{tikzpicture}
      \node[draw = none, fill = none] at (0,0){\includegraphics[trim={5 5 5 5}, clip, width=\width, height=\height]{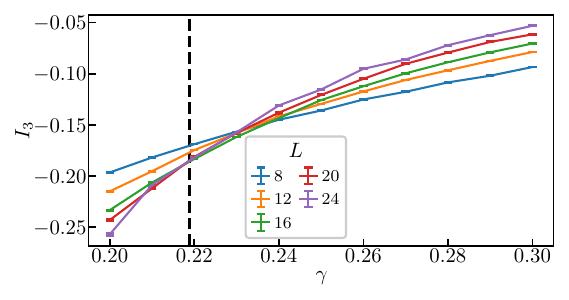}};
      \node[draw = none, fill = none] at (0.30*\width,-0.08*\height){\includegraphics[trim={5 5 5 5}, clip, width=\widthinset]{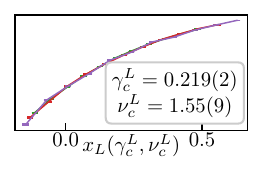}};
      
      \node[draw = none, fill = none] at (\width,0){\includegraphics[trim={5 5 5 5}, clip, width=\width, height=\height]{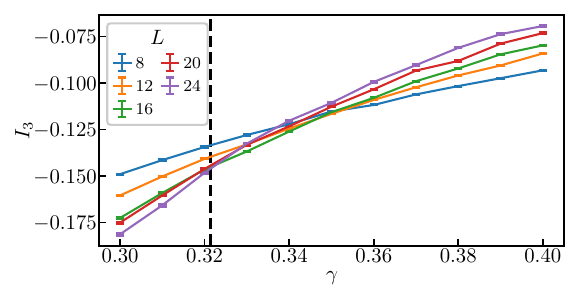}};
      \node[draw = none, fill = none] at (1.3*\width,-0.08*\height){\includegraphics[trim={5 5 5 5}, clip, width=\widthinset]{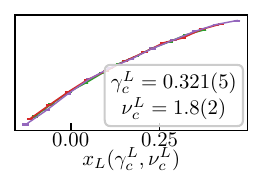}};
      
      \node[draw = none, fill = none] at (0,-\height){\includegraphics[trim={5 5 5 5}, clip, width=\width, height=\height]{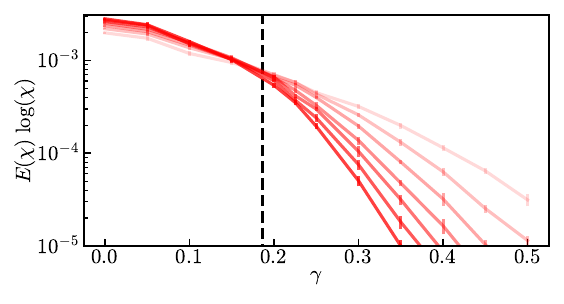}};
      \node[draw = none, fill = none] at (-0.16*\width,-1.07*\height){\includegraphics[trim={5 5 5 5}, clip, width=\widthinset]{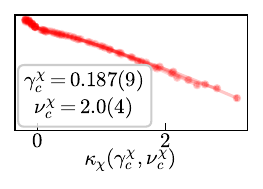}};
      
      \node[draw = none, fill = none] at (\width,-\height){\includegraphics[trim={5 5 5 5}, clip, width=\width, height=\height]{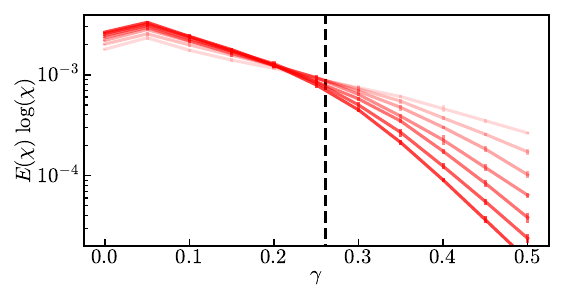}};
      \node[draw = none, fill = none] at (0.85*\width,-1.07*\height){\includegraphics[trim={5 5 5 5}, clip, width=\widthinset]{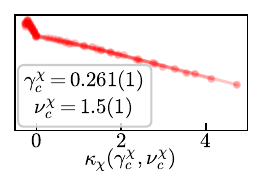}};
      
      \node[scale = \scale] at (-0.48*\width,0.5*\height) {\textbf{(a)}};
      \node[scale = \scale] at (0.56*\width,0.5*\height) {\textbf{(b)}};
      \node[scale = \scale] at (-0.48*\width,-0.49*\height) {\textbf{(c)}};
      \node[scale = \scale] at (0.56*\width,-0.49*\height) {\textbf{(d)}};

    \end{tikzpicture}
  }
  \caption{
    \textbf{Entanglement/Error phase transitions} in the XXX model, (a) and (c), and in the J-XXX model, (b) and (d).
    Dashed vertical lines correspond to the extracted critical measurement rates.
    (a),(b) Tripartite mutual information $I_3$ from exact diagonalization simulations, time-averaged for $t \in [2L, 4L]$, as a function of measurement rate $\gamma$ for several sizes $L$.  Inset: data collapse for $L \geqslant 16$.
    (c),(d) In semi-log scale, the projection error multiplied by $\log(\chi)$ obtained from TDVP algorithm as a function of measurement rate $\gamma$ for $\chi=n^2$ for $n\in[2,10]$ (from light to dark shades of red) with dashed vertical lines corresponding to $\gamma_c$ in the inset. Insets: data collapsed obtained from finite scaling analysis for $\chi=n^2$ for $n\in[4,10]$ (same colour code) with corresponding critical parameters with additional data points around transition. 
  }
  \label{fig_MIPT}
\end{figure*}

\textit{Models and methods.---}
We focus on two generic interacting systems in one dimension (whose MIPT has been also studied by different means also in \cite{PhysRevB.102.054302,Lumia2023,2308.09133})  namely a chain of $L$  spin-$1/2$ particles, unitarily evolving with $U(1)$ symmetric (magnetisation conserving) Hamiltonians, defined as 
\begin{equation}
\begin{aligned}
      \hat{H}_\textnormal{J-XXX} & =  \sum_{i=1}^L \left( \hat{S}^x_{i} \hat{S}^x_{i+1} + \hat{S}^y_{i} \hat{S}^y_{i+1} + \hat{S}^z_{i} \hat{S}^z_{i+1}  \right)  \\ 
      &  +  \sum_{i=1}^L J\left( \hat{S}^x_{i} \hat{S}^x_{i+2} + \hat{S}^y_{i} \hat{S}^y_{i+2}\right) 
\end{aligned}
    \ , 
\end{equation}
with spin operators $\hat{S}^\alpha_i$ and with $J=0$ (XXX chain) or $J=1/2$ (J-XXX chain). 
In addition to unitary evolution, the systems evolve under continuous weak monitoring of the local spin operator $\hat{S}_i^z$, for each site $i$ with a rate of measurement $\gamma$. The associated monitored dynamics of the quantum state are described by the
following stochastic Schrödinger equation (SSE) \cite{jacobs2006straightforward,wiseman2009quantum,barchielli2009quantum} for the many-body state $| \psi  \rangle$, 
\begin{multline}\label{eq:SSH} 
    d\ket{\psi_t}=-iHdt\ket{\psi_t} \\
    +\sum_{i=1}^{L}\left[ \sqrt{\gamma}(\hat{S}^z_i-\langle\hat{S}^z_i\rangle_t)dW_t^i-\frac{\gamma}{2}(\hat{S}^z_i-\langle\hat{S}^z_i\rangle_t)^2 dt \right]\ket{\psi_t}, 
\end{multline}
with the expectation value of the local magnetisation given by the state at a given time $\langle \hat{S}^z_i\rangle_t  = \langle \psi_t  | \hat{S}^z_i | \psi_t \rangle$.  Eq. \eqref{eq:SSH} can be easily simulated by 
alternating its unitary and measurement terms via a Trotter splitting, 
\begin{equation}
\label{eq:evol_state}
    \left|\psi_{t+\delta t}\right\rangle \approx C e^{\sum_{j=1}^L\left[\delta W_t^j+ 2\left\langle \hat{S}^z_j\right\rangle_t\gamma \delta t\right] \hat{S}^z_j} e^{-\mathbf{i} \hat{H} \delta t}\left|\psi_t\right\rangle  \ , 
\end{equation}
where the set of $\delta W^i$ are generated each time step from a normal distribution with variance $\sqrt{\gamma \delta t}$ and zero mean and $C$ is a normalizing constant.  As the measurements are only made by single site operators they do not require any TDVP projection on the space of MPS, differently from the unitary part, as we shall now describe.  We begin by expressing the wavefunction $\ket{\Psi({M})}$
as an MPS made up of the set of tensors $\{M\}$ each with a local basis $\left\{\left|\sigma_i\right\rangle\right\}_{\sigma_i=1}^{d_i}$ and bond dimensions $\{\chi_i\}$,
\begin{equation}
    |\Psi(M)\rangle=\sum_{\sigma_1, \ldots, \sigma_L}   M_{1; \chi_1}^{\sigma_1} \cdots M^{\sigma_L}_{L; \chi_{L-1}}\left|\sigma_1 \cdots \sigma_L\right\rangle
    \ ,
\end{equation}
where in the following we always refer to $\chi$ as the maximal value of the set $\{\chi_i\}$. Typically, time evolution of MPS is carried by methods such as TEBD \cite{PhysRevLett.91.147902}, where two or higher body gates are applied and entanglement is created between adjacent sites.  However, another approach is to fix a maximal bond dimension $\chi_i = \chi$ and then time evolve the MPS such that at each time step the evolution is projected back on the manifold of MPS with that given bond dimension.  This is achieved by introducing a projector $P_{\mathcal{T}_M\mathcal{M}}$ which projects onto the tangent space of the state $|\Psi(M)\rangle$ where fixed bond dimension $\chi$, at any time $t$, giving this way the TDVP time evolution
\begin{equation}\label{eq:TDVPeq}
    i \partial_t|\Psi(M)\rangle=P_{\mathcal{T}_M \mathcal{M}} \hat{H}|\Psi(M)\rangle \ .
\end{equation}
Projecting to the manifold of fixed bond dimension leads to a projection error  $| \phi \rangle  = \hat{H}\ket{\Psi(M)} -P_{\mathcal{T}_M \mathcal{M}} \hat{H}|\Psi(M)\rangle$, representing the orthogonal state to the local manifold of MPS. 
In our analysis, we define the norm of this residue as the projection error rate, and, given the projector $P_\psi(M_\chi)$ on the manifold of bond dimension $\chi$, this latter is given by 
\begin{equation}
        E(\chi)=|| \hat{H}  | \psi \rangle - P_\psi(M_\chi) \hat{H} | \psi \rangle  ||^2 ,
\end{equation}
which can be easily evaluated \cite{PhysRevB.88.075133,PhysRevB.97.045125,PhysRevB.94.165116} and where, most importantly, it can be decomposed in terms of $L$ \textit{local terms}. We remark that, since the projector depends on the state itself (in particular on two copies of the state), this expression is non-linear in the density matrix $\rho = \ket{\psi}\bra{\psi}$, therefore making it a good quantity to detect MIPT.

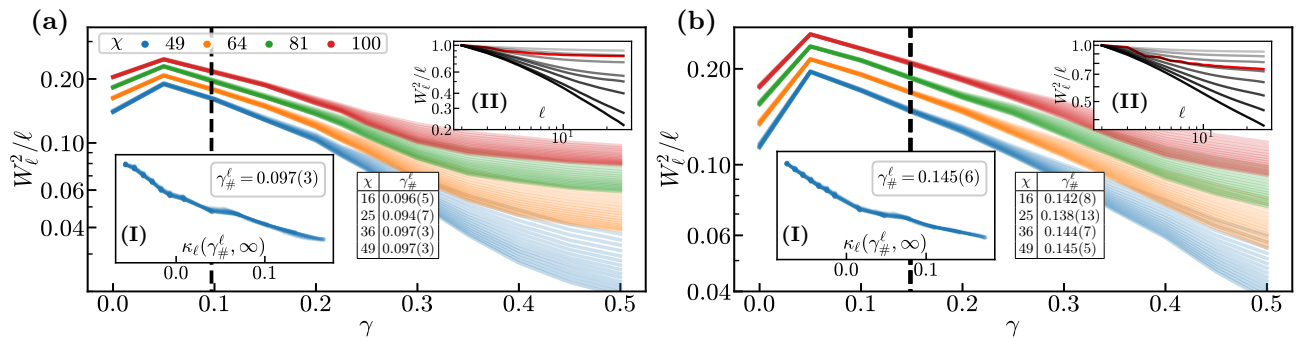
\begin{figure*}[t]
  \centering
  \resizebox{\textwidth}{!}{
    \def\width{4in}
    \def\widthinset{1.4in}
    \def\height{2in}
    \def\scale{1.3}
    \begin{tikzpicture}
      \node[draw = none, fill = none] at (0,0){\includegraphics[trim={5 5 5 5}, clip, width=\width, height=\height]{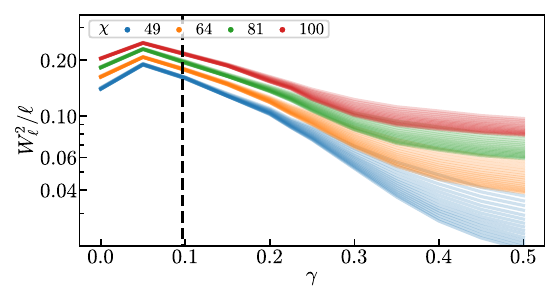}};
       \node[draw = none, fill = none] at (0.30*\width,0.28*\height){\includegraphics[trim={5 5 5 5}, clip, width=\widthinset]{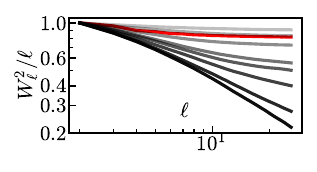}};
      \node[draw = none, fill = none] at (-0.16*\width,-0.11*\height){\includegraphics[trim={5 5 5 5}, clip, width=\widthinset]{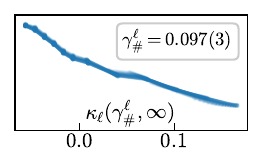}};
      
      \node[draw = none, fill = none] at (\width,0){\includegraphics[trim={5 5 5 5}, clip, width=\width, height=\height]{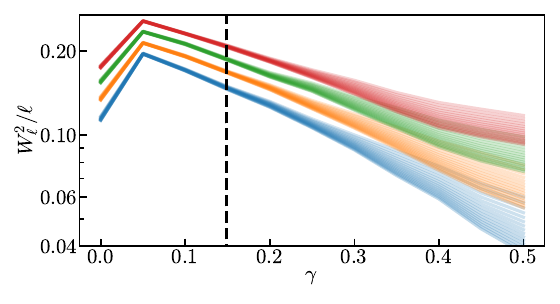}};
      \node[draw = none, fill = none] at (1.29*\width,0.28*\height){\includegraphics[trim={5 5 5 5}, clip, width=\widthinset]{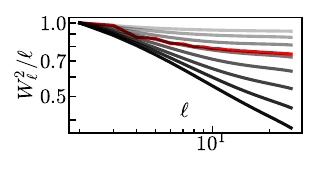}};
      \node[draw = none, fill = none] at (0.863*\width,-0.1*\height){\includegraphics[trim={5 5 5 5}, clip, width=\widthinset]{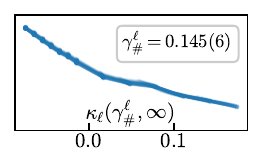}};

      \node[scale = \scale] at (-0.43*\width,0.5*\height) {\textbf{(a)}};
      \node[scale = \scale] at (0.57*\width,0.5*\height) {\textbf{(b)}};
      \node[scale = 0.8*\scale] at (-0.3*\width,-0.16*\height) {\textbf{(I)}};
      \node[scale = 0.8*\scale] at (0.25*\width,0.25*\height) {\textbf{(II)}};
      \node[scale = 0.8*\scale] at (0.723*\width,-0.16*\height) {\textbf{(I)}};
      \node[scale = 0.8*\scale] at (1.24*\width,0.25*\height) {\textbf{(II)}};

      \node [scale = 0.55*\scale,shape=rectangle,inner sep=0pt,fill=white](table1) at (1.1,-0.5) {
            \begin{tabular}{|c|c|} 
                \hline
                $\chi$ & $\gamma^{\ell}_{\#}$ \\ 
                 \hline
                16 & 0.096(5)  \\
                25 & 0.094(7) \\ 
                36 & 0.097(3) \\
                49 & 0.097(3) \\
                \hline
            \end{tabular}
        };

      \node [scale = 0.55*\scale,shape=rectangle,inner sep=0pt,fill=white](table1) at (11.5,-0.5) {
            \begin{tabular}{|c|c|} 
                \hline
                $\chi$ & $\gamma^{\ell}_{\#}$ \\ 
                 \hline
                16 & 0.142(8)  \\
                25 & 0.138(13) \\ 
                36 & 0.144(7) \\
                49 & 0.145(5) \\
                \hline
            \end{tabular}
        };
    \end{tikzpicture}
  }
  \caption{ 
    \textbf{CS phase transitions} for (a) the XXX model and (b) the J-XXX model.
    Dashed vertical lines correspond to the extracted critical measurement rates.
    Main plots: in semi-log scale, magnetization variance divided by $\ell$ shifted by 0.02 for each different $\chi$s (to increase readability) and for increasing $\ell\in[2,25]$ (light to dark shades of colours), dashed vertical line corresponds to $\gamma_{\#}$ from inset. Insets(I) data collapsed for $\chi=49$ with corresponding $\gamma_{\#}$. Insets(II), show in log-log scale, $W^2_{\ell}/\ell$ as a function of $\ell$ for $\gamma=0.05 \ n$ with $n\in [0,10]$ (light to dark shades of gray with $\gamma=0$ shown in red) for $\chi=100$. The inset tables show the extracted values of the critical transition rate (with KT scaling) at different values of $\chi$.   
  }
  \label{fig_charge_sharpening}
\end{figure*}

In order to benchmark the TDVP data, we also implement exact diagonalisation (ED) simulations, where one can probe the {tripartite mutual information} $I_3$ \cite{zabalo2020criticalpropertiesof} and detect the entanglement transition from the crossing of $I_3$ (obtained by partitioning the system into three parts of equal lengths) for different $L$s as function of $\gamma$. 
In order to find the correct critical parameters for the transitions, a finite-size scaling analysis of a given observable $\mathcal{O}$ is performed with
$ \mathcal{O} (\gamma, A) \sim f\left[x(\gamma, \gamma_\mathcal{O}, A, \nu_\mathcal{O}) \right]$, 
where $A$ corresponds to either $L$, in the case of ED simulations, or $\chi$ in the case of TDVP, and $x$ is an appropriate scaling function.
The standard {ansatz} for the scaling function is 
    $ x = x_A(\gamma_\mathcal{O}, \nu_\mathcal{O}) = (\gamma - \gamma_\mathcal{O}) A^{1/\nu_\mathcal{O}} $ together with the logarithmic (KT) scaling in $A$ in the limit $\nu_\mathcal{O} \rightarrow \infty$,
$ 
    x = x_A(\gamma_\mathcal{O}, \infty) = (\gamma - \gamma_\mathcal{O}) \log(A).
$
In order to fit the MPS transition, given the much quicker convergence in $\chi$ on the left of the critical point, see Fig \ref{fig_MIPT},  we need to introduce a \textit{piece-wise ansatz}, namely 
\begin{equation}
    x =  \kappa_A(\gamma_\mathcal{O}, \nu_\mathcal{O}) = 
    \left\{
    \begin{array}{lll}
        (\gamma - \gamma_\mathcal{O})  & \gamma < \gamma_\mathcal{O} \\
        (\gamma - \gamma_\mathcal{O}) A^{1/\nu_\mathcal{O}} &  \gamma > \gamma_\mathcal{O}
    \end{array}
    \right. \ ,
\end{equation}
and we cross-checked this scaling with the one where only the right side of the transition is fitted, giving analogous results.   When $\mathcal{O}(\gamma,A)$ probes the MIPT, then $\gamma_\mathcal{O} = \gamma_c^A$ and $\nu_\mathcal{O} = \nu_c^A$,
else if the CS is probed, $\gamma_\mathcal{O} = \gamma_\#^A$ and $\nu_\mathcal{O} = \nu_\#^A$. We refer to \cite{SM} for more details on the fitting procedure.

\textit{Results and discussions.---}
Starting always with the initial Néel state,
$  \ket{\Psi(0)} = \ket{ \downarrow \uparrow \downarrow \uparrow \dots \downarrow \uparrow} 
$ we perform monitored evolution at different values of $\chi$. In Fig. \ref{fig_time_evol} we show the time evolution of the error rate for two different values of $\gamma$ to illustrate how after a timescale of order $L$ the latter saturates to a steady value, which we plot as a function of $\chi$ in Fig \ref{fig_MIPT}.  As expected for the volume-law phase, at lower $\gamma$, $E(\chi) \log(\chi)$ converges to a constant value in $\chi$ at large $\chi$ for fixed $\gamma< \gamma_c$. By increasing $\gamma$, we observe a transition to a regime where the error decays exponentially with $\chi$, corresponding to the area-law phase. The existence of a transition can also be verified by ED simulations.
There, a crossing in $I_3$ as a function of $\gamma$ for different values of $L$ allows for the identification of the transition.
In both cases, the critical parameters are precisely found by a scaling analysis which reveals how MPS scaling determines a critical measurement rate $\gamma_c^\chi$ quite smaller than that of the ED, $\gamma_c^L$. Indeed, it is expected that the critical measurement rates obtained in a small system $L$ drift \textit{towards the left} with increasing $L$, as a smaller circuit is easier to volume-law entangle (which is visible also from Fig. \ref{fig_MIPT} (a,b)). The unitary (entangling) terms tend to dominate over measurements for small $L$, an effect which is indeed more prominent in the (longer range) J-XXX model.  The extracted $\gamma_c^\chi$ instead doesn't suffer from finite $L$ corrections as the system is here large enough ($L=60$ in all MPS simulations here), but it could in principle suffer from finite $\chi$ effects, which should instead cause a drift of the critical $\gamma$ \textit{towards the right}, as larger values of $\chi$ allow for more entanglement in the MPS. However, we find it very well-converged already for relatively small values of bond dimension. The two extracted critical exponents also slightly differ between ED and MPS, but in general, they may differ. Indeed, for pure states, a finite $\chi$ typically translates into a finite effective correlation length $L(\chi) \sim \chi^\kappa$  \cite{PhysRevLett.102.255701,
PhysRevB.78.024410,PhysRevLett.129.200601,PhysRevLett.123.250604}, with therefore a factor $\kappa$ between $L$-scaling and $\chi-$scaling. While the latter is indeed well established for equilibrium ground states, much less is known for non-equilibrium settings. 

Moving to the CS transition, we extract the late time value of the variance, and we plot it in Fig. \ref{fig_charge_sharpening}. For small $\gamma$, $W_\ell^2/\ell$ converges to a constant$(\gamma,\chi)$ (which, quite remarkably, is not a monotonous function $\gamma$) at large $\ell$, for fixed $\gamma < \gamma_\#$ and any $\chi$, meaning that the charge fluctuations are extensive in the sub-system size $\ell$. When $\gamma=0$ this scaling is expected by ETH-like arguments, as the system thermalises locally to a canonical state within the MPS manifold.
Increasing $\gamma$ enough, one encounters a transition to a phase where $W_\ell^2$ scales sub-linearly.   Once again, we observe that such behaviour converges very fast with bond dimension. Using a KT piece-wise scaling we can fix the critical $\gamma_\#$, which we find again slightly smaller than the ones predicted by ED simulations, see \cite{SM}, where the difference is again imputable to finite $L$ effects. Finally, we remark that our findings are only partially in agreement with the theory carried for a random circuit with large qudits \cite{PhysRevLett.129.120604}: despite the agreement on the predicted KT scaling, we find a transition from an extensive to a sub-extensive variance (as also observed in \cite{PhysRevB.107.014308} in small systems), where in \cite{PhysRevLett.129.120604} the transition is between an extended critical phase with $\log \ell$ scaling of the variance for $0<\gamma<\gamma_\#$ to a phase with sub-extensive scaling.  The differences could be imputable to either deviations from the theory, due to small/intermediate-scale physics, or to the TDVP effective classical dynamics used here. However, by cross-checking with ED simulations, we here provide strong evidence that the transition observed must be the same, even if the scaling of the variance in the two phases is apparently different. We shall leave this interesting question to further studies, yet we stress the importance of methods such as the one proposed here to probe the CS transition in large systems with continuous time.

\textit{Conclusion.---}
We have shown how TDVP time evolution of MPS  contains information on the MIPTs already at small values of bond dimensions. We have found a transition in the error rate of the TDVP, directly related to the entanglement transition. It is much easier to detect due to the dramatic change of its scaling with $\chi$ (exponential vs logarithmic, rather than logarithmic vs constant, for the entanglement entropy). Moreover, the advantage of using both the error rate and the charge fluctuations is that they can be expressed as a sum over \textit{local quantities}. One may understand the quick convergence of critical parameters with $\chi$ from the known fact that an ensemble of MPSs can successfully incorporate short (quantum-like) and long (classical-like) fluctuations, see \cite{white2009,PhysRevResearch.3.L022015}. As MIPTs probe the proprieties of the ensemble of quantum trajectories, it is reasonable that many of its features are already visible by an ensemble of MPS with relatively small bond dimensions, enough to incorporate the leading quantum effects. It remains therefore a lingering question to better clarify the scaling with $\chi$, as done for equilibrium critical states, and to employ our method, for example, to access MIPTs in higher dimensions. Finally, it will also be interesting to study more closely monitored TDVP evolution at small values of bond dimension as done for quantum scars dynamics  \cite{PhysRevX.10.011055,PRXQuantum.3.030343}

\textit{Acknowledgments.---}
We are extremely grateful to Xhek Turkeshi for enlightening discussions, help with different technical aspects and valuable feedback. We also thank Romain Vasseur, Adam Nahum, Andrea De Luca, Silvia Pappalardi, Michael Buchhold, Guglielmo Lami, Mario Collura, Luca Lumia, Laurens Vanderstraeten, Jutho Haegeman for  discussions. This work has been partially funded by the ERC Starting Grant 101042293 (HEPIQ) and the ANR-22-CPJ1-0021-01.  This work was granted access
to the HPC resources of IDRIS under the allocation
AD010513967R1 and A0140914149 and with a MPS code developed using the C++ iTensor library \cite{itensor-r0.3}. 

\bibliographystyle{apsrev4-2}
\bibliography{biblio}

\onecolumngrid
\newpage 
\appendix

\begin{center}
{\Large Supplementary Material \\ 
\titleinfo
}
\end{center}

\setcounter{equation}{0}
\setcounter{figure}{0}
\setcounter{table}{0}
\setcounter{page}{1}
\renewcommand{\theequation}{S\arabic{equation}}
\setcounter{figure}{0}
\renewcommand{\thefigure}{S\arabic{figure}}
\renewcommand{\thepage}{S\arabic{page}}
\renewcommand{\thesection}{S\arabic{section}}
\renewcommand{\thetable}{S\arabic{table}}
\makeatletter

\renewcommand{\thesection}{\arabic{section}}
\renewcommand{\thesubsection}{\thesection.\arabic{subsection}}
\renewcommand{\thesubsubsection}{\thesubsection.\arabic{subsubsection}}

\section{CS transition in exact diagonalization simulations}

To probe the CS transition, the initial state is chosen to be in a superposition of states in all possible charge sectors, 
\begin{equation}\label{eq:init_state_plus}
    \ket{\Psi(0)} = \bigotimes_{i = 1}^L \left(\ket{\uparrow} + \ket{\downarrow} \right) \ . 
\end{equation}

For the time evolution, we employ a first-order Trotterization protocol to separate the unitary evolution operator into layers of commuting 2-qubit gates.
In the case of the XXX model, the time-evolution operator for a single step is given by 
\begin{equation}
  \hat{U}_\textnormal{XXX} \approx \prod_{i} \hat{u}_{2i, 2i+1}^\textnormal{XXX} \prod_{i} \hat{u}_{2i-1, 2i}^\textnormal{XXX} \ , 
\end{equation}
with an odd followed by an even layer of 2-qubit gates
\begin{equation}
  \hat{u}_{i, i+1}^\textnormal{XXX} = \exp\left[- i \delta t \left( \hat{S}^x_{i} \hat{S}^x_{i+1} + \hat{S}^y_{i} \hat{S}^y_{i+1} + \hat{S}^z_{i} \hat{S}^z_{i+1}  \right)  \right] \ ,
\end{equation}
where $\delta t \ll 1$.
In the case of the J-XXX model, we apply the same evolution operator of the XXX model followed by a dense arrangement in three layers of all the possible 2-qubit gates
\begin{equation}
  \hat{u}_{i, i+2}^\textnormal{J} = \exp\left[- i \delta t   J\left( \hat{S}^x_{i} \hat{S}^x_{i+2} + \hat{S}^y_{i} \hat{S}^y_{i+2}\right) \right]\ .
\end{equation}

To implement the measurements, a strong measurement procedure was employed.
With probability $p = \gamma \delta t$, each site $i$ is measured according to the Born rule,
\begin{equation}
    \ket{\psi} \rightarrow P_{\pm} \ket{\psi} \quad , \quad P_\pm = \left( \frac{\mathds{1}}{2} \pm \hat{S}_i^z  \right) \ , 
\end{equation}
where $P_\pm$ is the projector or Kraus operator corresponding to $\hat{S}_i^z$ and $+$ or $-$ is chosen with corresponding probability $p_\pm = \bra{\psi} P_\pm \ket{\psi}$.

\begin{figure*}[b]
  \centering
  \resizebox{\textwidth}{!}{
    \def\width{4in}
    \def\widthinset{1.4in}
    \def\height{2in}
    \def\scale{1.3}
    \begin{tikzpicture}
      \node[draw = none, fill = none] at (0,0){\includegraphics[trim={5 5 5 5}, clip, width=\width, height=\height]{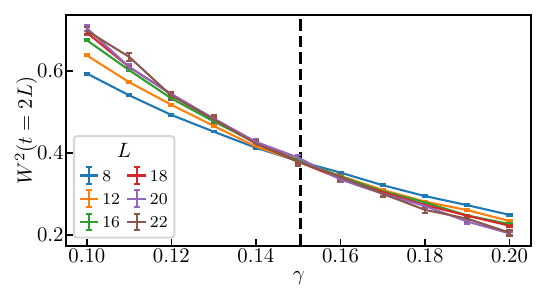}};
      \node[draw = none, fill = none] at (0.30*\width,0.22*\height){\includegraphics[trim={5 5 5 5}, clip, width=\widthinset]{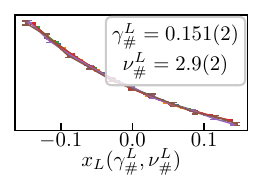}};
      
      \node[draw = none, fill = none] at (\width,0){\includegraphics[trim={5 5 5 5}, clip, width=\width, height=\height]{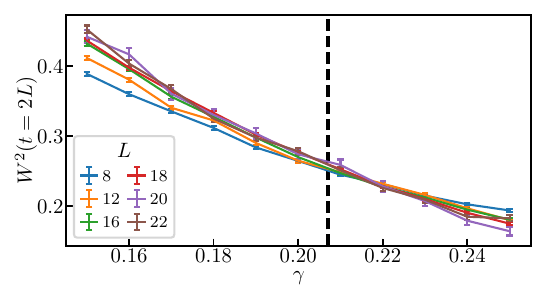}};
      \node[draw = none, fill = none] at (1.29*\width,0.24*\height){\includegraphics[trim={5 5 5 5}, clip, width=\widthinset]{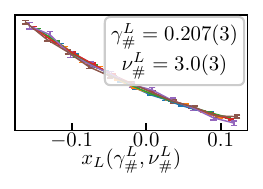}};

      \node[scale = \scale] at (-0.43*\width,0.4*\height) {\textbf{(a)}};
      \node[scale = \scale] at (0.57*\width,0.4*\height) {\textbf{(b)}};
    
    \end{tikzpicture}
  }
  \caption{ 
    Magnetization variance $W^2$ at time $t = 2L$ for (a) the XXX model and (b) the J-XXX model, as a function of $\gamma$ for different $L$.
    Dashed vertical lines correspond to critical measurement rates.
    Insets: data collapse and corresponding critical parameters.
  }
  \label{fig_charge_sharpening_ed}
\end{figure*}

Apart from the initial state, the simulations of Figs. \ref{fig_MIPT}(a,b) follow the same protocol. However, since the models are $U(1)$ conserving, meaning all the 2-qubit gates are block diagonalized as 
\begin{equation}
    \hat{u} = 
    \begin{pmatrix}
        \hat{u}_{1 \times 1} & & \\
        & \hat{u}_{2 \times 2} & \\
        & & \hat{u}_{1 \times 1} \\
    \end{pmatrix}
    \ ,
\end{equation}
if the initial conditions are such that we are bound to a single sector of the Hilbert space, then we need not simulate the other sectors.
Taking these symmetry considerations into account is important to perform more efficient simulations and reach larger system sizes and sampling pools \cite{10.1063/1.3518900}.

The measurements will eventually collapse the state of Eq. \ref{eq:init_state_plus} into a single sector, which can happen at different time scales, depending on $\gamma$.
For $\gamma < \gamma_\#$, the timescale is $\mathcal{O}(L)$ whereas for $\gamma > \gamma_\#$ it is sublinear in system size \cite{agrawal2022entanglmentandchargesharpening}. 
At any time, the spread of the active sectors is given by the variance
\begin{equation}
  W^2(t) = \langle Q^2(t) \rangle - \langle Q(t) \rangle^2 \ ,
\end{equation}
where, for the total magnetization of the system, we drop the label $\ell$ in Eq. \eqref{eq:Well}, meaning $\ell = L$. 

Due to the change in CS timescales, a crossing of $W^2$ at $t \propto L$ is observed at the critical measurement rate, Fig. \ref{fig_charge_sharpening_ed}.
Analogously to the main text we find that $\gamma_\#^L = 0.151(2)$ for the XXX model in Fig. \ref{fig_charge_sharpening_ed}(a) is smaller than that of the J-XXX, $\gamma_\#^L = 0.207(3)$ in Fig. \ref{fig_charge_sharpening_ed}(b).
These critical measurement rates are larger than corresponding ones found with the TDVP method, Fig. \ref{fig_charge_sharpening}, reflecting the phenomenology already found in the entanglement/error transition.

\section{Description of the fitting algorithm}

The correct critical parameters for the scaling functions are the ones that minimize the cost function \cite{doi:10.1143/JPSJ.62.435, zabalo2020criticalpropertiesof, PhysRevB.108.L020306}
\begin{equation}
    W( \gamma_\mathcal{O}, \nu_\mathcal{O}) = \frac{1}{n-2} \sum_{i = 2}^{n-1} w (x_i, y_i, d_i | x_{i-1}, y_{i-1}, d_{i-1}, x_{i+1}, y_{i+1}, d_{i+1}) \ , 
\end{equation}
where $x_i = x(\gamma_i, \gamma_\mathcal{O}, A_i, \nu_\mathcal{O}^A)$ is the scaling function value of the $i$th data point of a total of $n$ data points, sorted such that $x_{i+1} \geqslant x_i$, with values $y_i = \mathcal{O}(\gamma_i, A_i)$ and corresponding errors $d_i$.
The cost density is given by $w(x_i, y_i, d_i|\dots) = (y_i - \overline{y})^2/\Delta$, with 
\begin{equation*}
\begin{gathered}
    \overline{y} = \frac{(x_{i+1} - x_i)y_{i-1} - (x_{i-1} -x_i) y_{i+1}}{x_{i+1} - x_{i-1}} \\
    \Delta = d_i^2 + \left(\frac{x_{i+1}-x_{i}}{x_{i+1}-x_{i-1}}d_{i-1}\right)^2 + \left(\frac{x_{i-1}-x_{i}}{x_{i+1}-x_{i-1}}d_{i+1}\right)^2
\end{gathered}
\ .
\end{equation*}
If $x_{i-1} = x_{i+1}$ (which happens with the $\kappa_A$ scaling function), we get undefined $0/0$ terms in the cost density, which we resolve by taking the limit where $x_i$ sits precisely in the midpoint between $x_{i-1}$ and $x_{i+1}$ as they approach each other.
This cost function works by minimizing the distance from $y_i$ to the line determined by the points at $i-1$ and $i+1$, consequently locally aligning the points.

      

      
    

\end{document}